\documentclass[twocolumn]{article}
\usepackage{geometry}
\usepackage{graphicx}
\usepackage{bbm}
\usepackage[utf8]{inputenc}
\usepackage[T2A,T1]{fontenc} 
\usepackage{amssymb}
\usepackage{amsmath}
\usepackage{graphics}
\usepackage{times}
\usepackage{bbm}
\usepackage{tabularx}
 
\usepackage{psfrag} 
\usepackage{eurosym}
\usepackage{titletoc}
\usepackage{esint}
\usepackage{multicol}
\def\€{\euro{}}
\usepackage{url}
\usepackage{color}
\usepackage{hyperref}
\usepackage{array}
\usepackage[normalem]{ulem}
\usepackage{float}
\definecolor{linkcolor}{rgb}{0,0,0.6} % définition de la couleur des liens pdf  
\usepackage{fancyhdr} %en-tête de page personnalisés 
\hypersetup{ 
    colorlinks, 
    citecolor=black, 
    filecolor=black, 
    linkcolor=black, 
    urlcolor=linkcolor 
}   
\DeclareGraphicsExtensions{.jpg, .eps}
\usepackage{etoolbox}
\usepackage[sort&compress,comma]{natbib}
\bibpunct{[}{]}{,}{n}{,}{,}
\usepackage{wrapfig}
\usepackage[detect-all]{siunitx}

\begin{document}

\title{Transmission XMCD-PEEM imaging of a vertical FEBID cobalt nanowire with engineered geometry}

%\maketitle{}

\begin{center}
\textbf{\Large{Transmission XMCD-PEEM imaging of an engineered vertical FEBID cobalt nanowire with a domain wall}}
\end{center}

\begin{center}
Alexis Wartelle$^{1}$,
Javier Pablo-Navarro$^{2}$,
Michal Sta\v{n}o$^{1}$,
Sebastian Bochmann$^{3}$,
S\'{e}bastien Pairis$^{1}$,
Maxime Rioult$^{4}$, 
Christophe Thirion$^{1}$,
Rachid Belkhou$^{4}$,
Jos\'{e} Mar\'{i}a de Teresa$^{2,5,6}$,
C\'{e}sar Mag\'{e}n$^{2,6,7}$ and
Olivier Fruchart$^{1,8}$
\end{center}

\footnotesize{
\noindent$^{1}$ Univ. Grenoble Alpes, CNRS, NEEL, F-38000 Grenoble, France \\
$^{2}$ Laboratorio de Microscop\'{i}as Avanzadas (LMA), Instituto de Nanociencia de Arag\'{o}n (INA), Universidad de Zaragoza, E-50018 Zaragoza, Spain \\
$^{3}$ Friedrich-Alexander-Universit\"{a}t, Erlangen, Germany\\
$^{4}$ Synchrotron-SOLEIL, Saint-Aubin, BP48, F91192 Gif sur Yvette Cedex, France \\
$^{5}$ Instituto de Ciencia de Materiales de Arag\'{o}n (ICMA), Universidad de Zaragoza-CSIC, E-50009 Zaragoza, Spain \\
$^{6}$ Departamento de F\'{i}sica de la Materia Condensada, Universidad de Zaragoza, E-50009 Zaragoza, Spain \\
$^{7}$ Fundaci\'{o}n ARAID, 50018 Zaragoza, Spain \\
$^{8}$ Univ. Grenoble Alpes, CNRS, CEA, Grenoble INP\footnote{Institute of Engineering Univ. Grenoble Alpes}, SPINTEC, F-38000 Grenoble, France
}
%\begin{tabularx}{\textwidth}{lX|lX}
%$^{1}$ & Univ. Grenoble Alpes, Inst NEEL, F-38000 Grenoble, France & $^{5}$ & Friedrich-Alexander-Universit\"{a}t, Erlangen, Germany\\
%$^{2}$ & CNRS, Inst NEEL, F-38042 Grenoble, France & $^{6}$ & Synchrotron-SOLEIL, Saint-Aubin, BP48, F91192 Gif sur Yvette Cedex, France \\
%$^{3}$ & Departamento de F\'{i}sica de la Materia Condensada, Universidad de Zaragoza, Facultad de Ciencias, Zaragoza 50009, Spain & $^{7}$ & CNRS, SPINTEC, F-38000 Grenoble, France \\
%$^{4}$ & Instituto de Ciencia de Materiales de Arag\'{o}n, Universidad de Zaragoza-CSIC, Facultad de Ciencias, Zaragoza 50009, Spain  & $^{8}$ & CEA, INAC-SPINTEC, F-38000 Grenoble, France \\
%\end{tabularx}

\begin{center}
\normalsize{Corresponding author: A. Wartelle (email: alexis.wartelle@neel.cnrs.fr).}
\end{center}

\textbf{
Using focused electron-beam-induced deposition (FEBID), we fabricate vertical, platinum-coated cobalt nanowires with a controlled three-dimensional structure. The latter is engineered to feature bends along the height: these are used as pinning sites for domain walls, the presence of which we investigate using X-ray Magnetic Circular Dichroism (XMCD) coupled to PhotoEmission Electron Microscopy (PEEM). The vertical geometry of our sample combined with the low incidence of the X-ray beam produce an extended wire shadow which we use to recover the wire's magnetic configuration. In this transmission configuration, the whole sample volume is probed, thus circumventing the limitation of PEEM to surfaces. This article reports on the first study of magnetic nanostructures standing perpendicular to the substrate with XMCD-PEEM. The use of this technique in shadow mode enabled us to confirm the presence of a domain wall (DW) without direct imaging of the nanowire.
}

%\strut
%
%\begin{center}
%\emph{Keywords:} \textbf{FEBID, ferromagnetic nanowires, three-dimensional nanostructures, XMCD-PEEM, transmission microscopy, core-shell structures, domain walls}
%\end{center}

\begin{figure*}[t!]
\centering\includegraphics[width=0.99\textwidth]{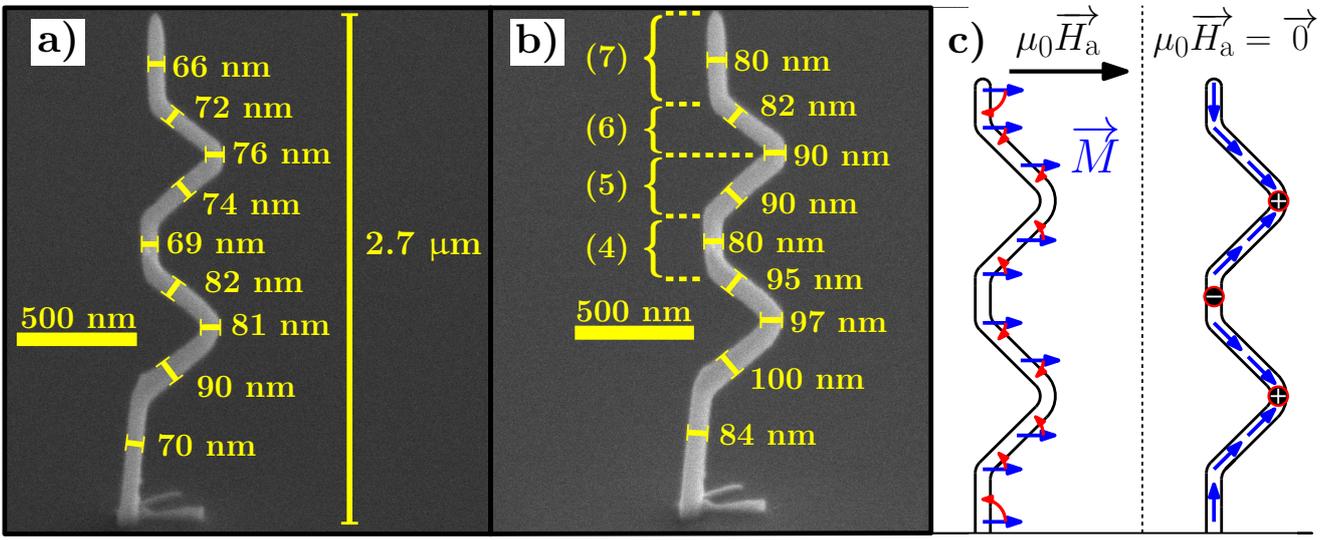}
\caption{SEM views at \SI{52}{\degree} stage tilt (with diameter and tilt-corrected height indications) of a) the as-grown Co FEBID nanowire and b) the structure after FEBID Pt coating. The four top segments (4), (5), (6) and (7) will be discussed further in the section pertaining to the magnetic state of the nanowire. c) Scheme for application of the in-plane magnetic field leading to three DWs of alternating polarity (red-rimmed black dots) at remanence.}
\label{fig_sem_nw_asgrown}
\end{figure*}

%\begin{wrapfigure}{r}[0cm]{0.45\textwidth}
%\centering\includegraphics[width=0.449985\textwidth]{fig_F11_NW1_red_with_fieldscheme_v2.pdf}
%\caption{a) SEM view of the as-grown Co FEBID nanowire F1 with indications of dimensions in yellow b) Scheme for application of the in-plane field leading to two DWs at remanence.}
%\label{fig_sem_nw_asgrown}
%\end{wrapfigure}
\normalsize{

\section{Introduction}

%\begin{onecolumn}

As nanofabrication techniques and magnetic imaging methods evolve, novel magnetic nanostructures not limited to two dimensions are increasingly investigated \cite{kimling_peem_coreshell,araujo2015_CrossedPCmembranesAndNi-NiFenanowires}. The use of bottom-up approaches such as electrodeposition into nanoporous templates \cite{iglesias-freire2015_DiameterModulatedFeCoCu_NWs,lavin2012_conws_angdep_coercivityeltit}, strain engineering \cite{streubel2013_RolledSingleLayerMembranes} and self-assembly induced by the decomposition of a thin-film matrix \cite{mohaddes-ardabili2004_SelfAssemblyOfSingleCrystalFeNWsFromDecompositionOfMatrix} has enabled the production of magnetic micro- or nanoparticles with a large variety of geometries and materials. However, the samples are often limited to straight cylindrical or tubular shapes \cite{fukunaka2006_ElectrodepositionOfNWsAndNTsSeveralMetals,li2014_NiNTarrayInPCmembraneElectrolessDeposition,Richardson10032015,schaefer2016_electrolessNiCoNTs,Zhang2013_ElectrodepositionOfNTs}, a restriction that can be lifted with FEBID. This technique allows one to deposit inside a Scanning Electron Microscope (SEM) the desired material on the electron beam spot with a lateral size below \SI{100}{\nano\meter}  \cite{deteresa_review_febid_nanostructures,deTeresa2016_MagneticNanostructuresGrownByFEBID} with the possibility to move that spot during growth as well as rotate or tilt the substrate. As a result, it is possible to grow structures in two or three dimensions \cite{gavagnin_nanopillars_mfm_febid_needle,fernandezpacheco2013_3DnanowiresByMOKE,fowlkes2016_3DnanomanufacturingWithFEBID} with the few limitations arising from the directional character of the technique. Moreover, tuning the purity of the deposited metal either during the deposition process \cite{Serrano-Ramon2011_ACSnano_FEBIDgrownFerromagneticNanostructures,shawrav2016_HighPurityFEBIDgold} or post-fabrication \cite{szkudlarek2015_AnnealingOfFEBIDcopperMatGrowthOfPureCuNanoXtals} allows one to further tailor the structure properties. In the case of magnetic nanostructures, cobalt is most often used \cite{deteresa_review_febid_nanostructures}, and the deposits are polycrystalline with no texture and grain size below \SI{10}{\nano\meter} \cite{cordoba2011_InvestigationOfCoFEBIDstructuresWithSTEM-EELSandHRTEM}.

\begin{figure*}[t]
\centering\includegraphics[width=0.9\textwidth]{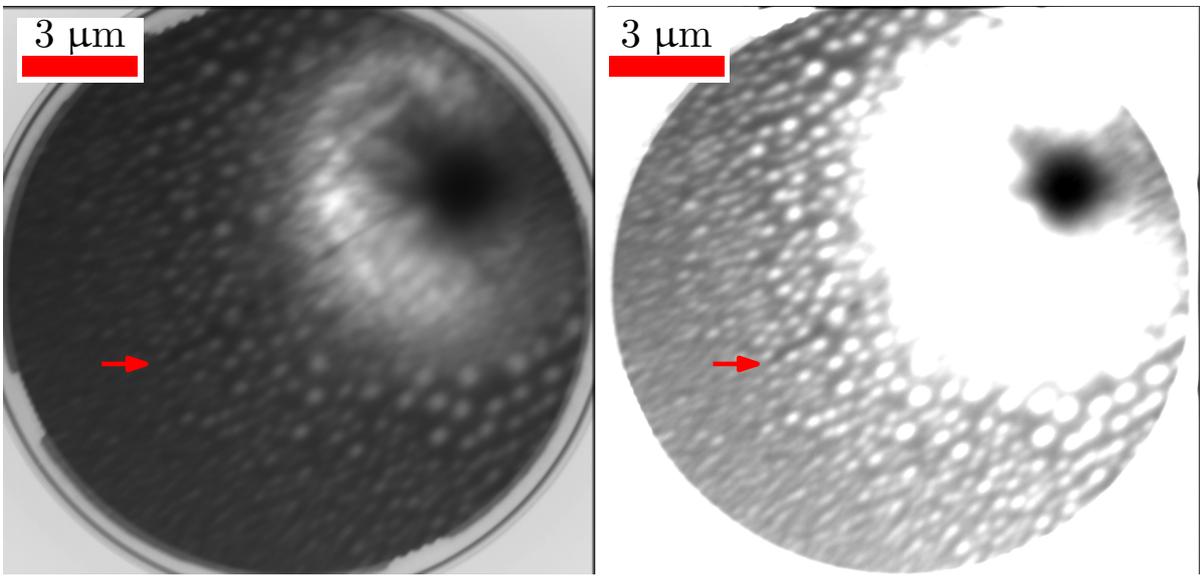}
\caption{PEEM images of the surroundings of the nanostructure. The two images differ only in the linear gray scale: the range is 4.3 times smaller on the right with a slightly higher minimum, so that the maximum extent of the nanostructure's shadow (whose direction is that of the X-ray beam) is better seen. On both images, the latter's tip is indicated by the red arrow.}\label{fig_cut_shadow}
\end{figure*}

We investigate here Co structures grown in the shape of near-cylindrical nanowires perpendicular to their silicon wafer substrate and coated with FEBID-grown Pt \cite{pablo-navarro2016_CoreShellFEBIDcoPtNWs}, as explained in the following section. They were fabricated with several well-defined bends as can be seen in Fig.\ref{fig_sem_nw_asgrown}.a) and b), featuring the nanostructure which will be the focus in the following. The purpose of its geometry is to obtain DWs at remanence as sketched in Fig.\ref{fig_sem_nw_asgrown}.c), after having saturated the sample with a horizontal magnetic field (i.e. in the plane of the substrate) aligned with the bends. Such a method was demonstrated in soft cylindrical nanowires \cite{dacol2016_nucleation_imaging_motion_dws} and is also applicable in our case considering the soft nature of our deposited magnetic material. Since magneto-optical techniques and magnetic force microscopy are not suitable for such samples, we use shadow XMCD-PEEM \cite{kimling_peem_coreshell,streubel2012_HemisphericalPyCapsShadowXMCDPEEM,dacol_bp_id,jamet2015_quantitative_xmcd} to determine the magnetic state of the structure and assess the presence of DWs. With respect to experiments involving nanostructures lying on the substrate \cite{Bran2016_XMCDPEEMonBambooNWsFeCoCu,dacol_bp_id,streubel2012_HemisphericalPyCapsShadowXMCDPEEM}, the vertical geometry provides access to a larger proportion of the shadow. Furthermore, the spectroscopic capabilities of the technique allow us to probe the metallic nature of the deposited material in the volume thanks to the transmission of X-rays through the sample. To our knowledge, this is the first study of such vertical elongated nanostructures with shadow XMCD-PEEM. It demonstrates the technical feasibility of imaging this type of nanowires and paves the way for the investigation of more complex vertical structures such as core-shell nanowires with different ferromagnetic materials. Indeed, shadow XMCD-PEEM probes magnetization in the sample volume with the added value of element-specificity, whereas e.g. electron holography \cite{phatak2014_CoNanospiralsWithElectronHolography} is only sensitive to the total induction.

%\end{onecolumn}

\section{Sample fabrication}

The three-dimensional (3D) core-shell structure was grown in the commercial Helios Nanolab 650 Dual Beam system using the Schottky field-emission gun (S-FEG) electron column and the gas injector systems (GIS) for Co and Pt depositions using Co$_2$(CO)$_8$ and CH$_3$CpPt(CH$_3$)$_3$ precursor gases, respectively. The substrate was a Si wafer with native oxide. Firstly, the Co nanowire core was grown with a \SI{5}{\kilo\volt} electron voltage and a \SI{100}{\pico\ampere} beam current at a gas pressure of \SI{3.33 e-5}{\milli\bar} (base pressure of \SI{4.73 e-6}{\milli\bar}). The stage stands still with zero degree tilt throughout the Co deposition. The pattern is composed of 77 points separated \SI{14}{\nano\meter} in a straight line parallel to the flat end of the Co GIS, keeping constant the precursor molecules flux over the whole deposition. The nanowire is made of seven segments, each one having its particular growth strategy. Vertical segments are obtained by scanning a single pattern point while the electron beam stands still; by contrast, bent segments are fabricated by scanning a sequence of points while shifting the electron beam position. For a fixed total horizontal shift during the sequence is fixed, the segment's angle with respect to the substrate depends on the number of points. The higher that number, the shorter the distance between two subsequent points: therefore the overlap between them will be higher and the angle with respect to the substrate will increase. In our case, each bent segment corresponds to 18 pattern points with a scanning time of \SI{97.2}{\milli\second}. To form the \SI{90}{\degree} bends, the joint between the bent segments is fabricated by a single point scanned for \SI{581.2}{\milli\second}. Then, the bend is completed by reversing the electron beam shift direction.

The first segment of the nanowire core was grown by depositing on the first point (\SI{2903.8}{\milli\second} at the same position). The second segment and the third one were then grown as described above, taking into account that the points of the latter are exactly over those of the former but gone through in the reverse direction, thus forming the first bend of the nanostructure. The forth segment growth was completed within \SI{1549.1}{\milli\second} scanning of the same point. Then, the fifth and the sixth segments, as well as the joint between them were carried out just as the first bend. The top segment was fabricated scanning the last point during \SI{2419.8}{\milli\second}. The Pt shell was grown immediately after the ferromagnetic core ($\sim$65-70\% at. Co) in order to avoid its oxidation, following the process described in a previous work \cite{pablo-navarro2016_CoreShellFEBIDcoPtNWs}. The experimental conditions were \SI{5}{\kilo\volt} electron voltage and \SI{100}{\pico\ampere} beam current at \SI{2.40 e-5}{\milli\bar}. First, tilting the stage \SI{52}{\degree} the nanostructure is seen as in Fig.\ref{fig_sem_nw_asgrown}.a) and a polygonal pattern of Pt deposition (\SI{2}{\second}) is set following the shape of the nanowire core viewed from this perspective. Then, to complete the coating, the stage is rotated \SI{180}{\degree} and a new deposit is carried out in the same way just on the opposite side. Pt deposition increases the nanowire diameter by approximately \SI{13}{\nano\meter}, as seen in Fig.\ref{fig_sem_nw_asgrown}.b), just enough to prevent Co oxidation.

\section{XMCD-PEEM imaging and characterization}\label{xmcd-peem_imaging}

The XMCD-PEEM experiments have been performed on the XPEEM branch of the HERMES beamline (Synchrotron SOLEIL - France) \cite{belkhou2015_XRmicroscopyAtHermes}. Several silicon wafers featuring rows of nanostructures similar to the aforementioned sample were brought for investigation at the Co L$_3$ edge, which is associated to large magnetic dichroism. Despite the apparent fragility of the nanowires arising from their high aspect ratio, the majority of those which we tried to image on site were still standing. Aside from shocks during transportation, the structures may be damaged during sample mounting and also by the imaging technique itself. Indeed, as the samples are subjected to a high voltage of \SI{20}{\kilo\volt} (with respect to the microscope objective), a large field emission of electrons due to the tip of the structures is generated. The associated current, as well as the resonant absorption of X-rays, can result in significant overheating since the only heat sink in the vacuum environment is the Si substrate through the narrow base of the structure.

This field emission is however of practical use in the search for standing structures as it creates wide, strongly contrasted features directly at the location of the nanowires. It corresponds to the dark area in the top right quadrants of the PEEM images in Fig.\ref{fig_cut_shadow}. The low intensity is expected; indeed, the current associated to the field-emission does not contribute to the image because it is rejected by the energy slits of the microscope. In addition, this positive current flowing downwards i.e. along the structure towards its foot, results in a negative charge density close to the wire's tip, as well as an Oersted-type magnetic field. Both deflect the photoelectrons from the nanostructure, which would build up the direct image on the detector. Since these electrons are selected with a contrast aperture rejecting photoelectrons with a wave-vector different from that of the photoelectrons from the substrate, only a vanishing signal is detected in this area. Furthermore, distortions due to the strong and inhomogeneous electric fields at the nanowire tip must be taken into account. As a result, no direct imaging of the structure could be performed and hence no XMCD-PEEM imaging of the nanowire's surface magnetization. However, when visible, its shadow (resulting from resonant absorption of X-rays) contains enough information \cite{kimling_peem_coreshell} to deduce the structure's magnetic state, as will be shown later on.

Due to the incidence angle of \SI{18}{\degree}, the total shadow length is increased with respect to the structure height by a factor $1/\sin{(\SI{18}{\degree})}\simeq3.2$. Therefore, one way to retrieve the nanowire's height is to measure the extent of its shadow: it can easily be traced on both sides since the center of the very dark area on the PEEM images corresponds to the location of the structure. From the images, the shadow length is about \SI{8.1\pm0.3}{\micro\meter}, which results in a height of \SI{2.5\pm0.1}{\micro\meter}, while SEM measurements performed before shipping of the samples indicated \SI{2.7}{\micro\meter}. SEM investigation after the synchrotron experiments indicated that the nanostructure was bent with respect to its initial state, as is illustrated in Fig.\ref{fig_f11_1_postmeas}. The measured height was \SI{2.3\pm0.1}{\micro\meter}.
%!!!!!!!!!!!!!!!!!!!!!!!!!!!!!!!!!!!!!!!!!!!!!!!!!!!!!!!!!!!!!!!!!!!!!!!!!!!!!!!!!!!!!!!!!!!!!!!!!!!!!!!!!!!!!!!!!!!!!!!!!!!!!!!!!!!!!!!!!!!!!!!!!!!!!!!!!!!!!!!!!!!%\textcolor{red}{check with S\'{e}bastien how lengths are with tilt of 45 deg} !!!!!!!!!!!!!!!!!!!!!!!!!!!!!!!!!!!!!!!!!!!!!!!!!!!!!!!!!!!!!!!!!!!!!!!!!!!!!!!!!!!!!!!!!!!!!!!!!!!!!!!!!!!!!!!!!!!!!!!!!!!!!!!!!!!!!!!!!!!!!!!!!!!!!!!
The agreement between both height estimates confirms that some plastic deformations were induced, possibly during observation. Such deformations have previously been observed for such structures simply upon aging. Since in this case the sample was also exposed to (resonantly absorbed) X-rays as well as large electric fields, such differences with respect to its initial state are not surprising.

%Since its first experimental evidence \cite{schuetz1987_FirstExpEvidenceOfXMCDinFe} about thirty years ago, XMCD has been used as a powerful asset in microscopic investigation of magnetic materials and nanostructures \cite{cheng_keavney_xpeem_review}. Its combination with PEEM allows for microscopic investigation of magnetization patterns in metallic samples \cite{pizzini_vogel_fruchart_2012,kimling_peem_coreshell,Choe2004_VortexGyrotropicMotionWithXMCD-PEEM,Nolting2000_InterfacialCouplingRevealedWithXMCDPEEM} with a typical resolution of a few tens of nanometers \cite{locatelli2006_speleemAtElettra,belkhou2015_XRmicroscopyAtHermes}. Since PEEM is a surface technique with a limitation of probed depth of about $\SI{10}{\nano\meter}$ (corresponding to the photoelectron's mean free path inside matter), conventional XMCD-PEEM probes magnetization on the surface for samples that are not ultra-thin films. As soon as the samples are not prepared in-situ, the possibility of oxidation arises: the latter may decrease the dichroic signal as the contribution of a non-ferromagnetic oxide shell increases with respect to the ferromagnetic metal.

\begin{figure}[!h]
\centering\includegraphics[width=0.48\textwidth]{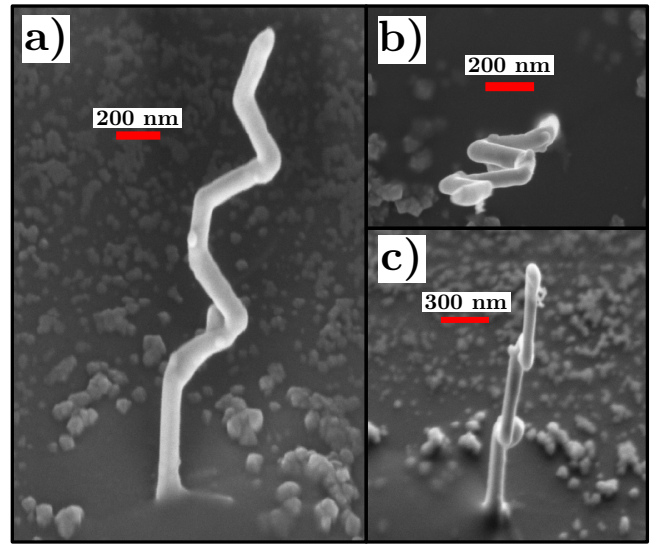}
\caption{SEM observations of the nanowire after the XMCD-PEEM experiments a) from the top, untilted stage b) along the bends with \SI{45}{\degree} tilted stage c) still at \SI{45}{\degree} tilt, view perpendicular to the bends.}\label{fig_f11_1_postmeas}
\end{figure}

\begin{figure*}[t]
\centering\includegraphics[width=0.98\textwidth]{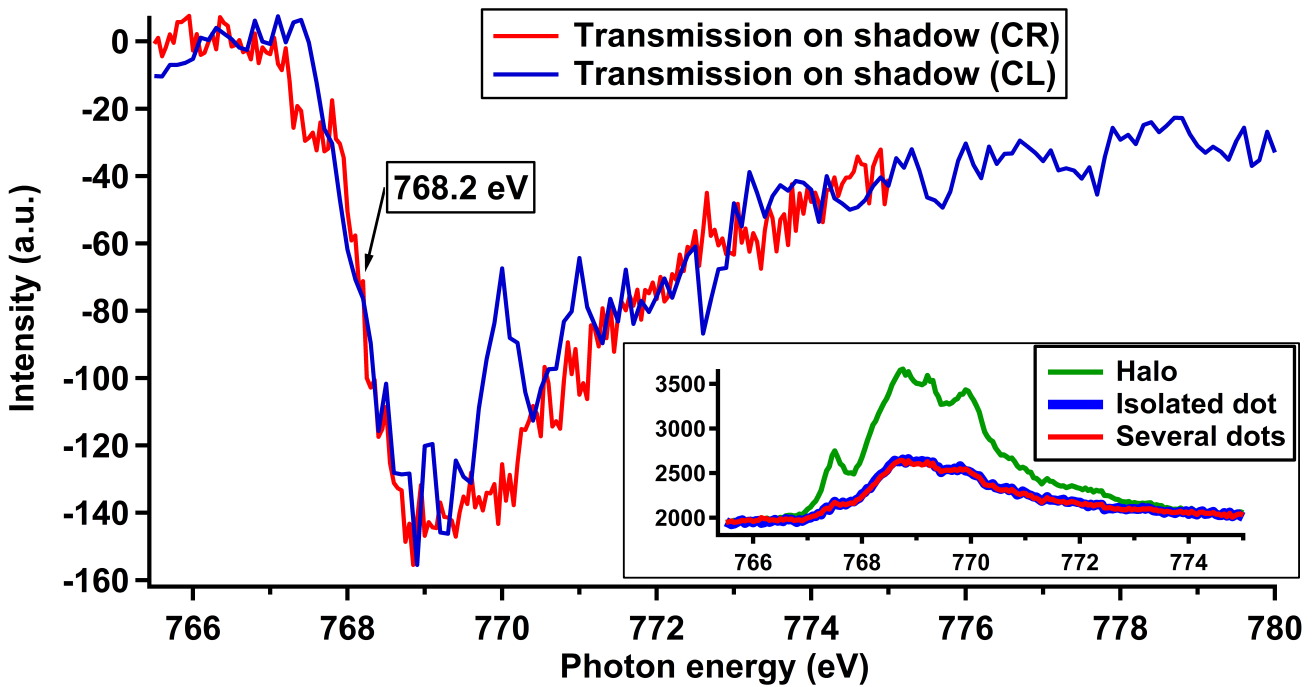}
\caption{XAS on the structure's shadow, featuring the decrease in transmission as the photon energy is sweeped through the Co L$_3$ edge. The blue (red) curve corresponds to circularly left (right) polarized light. The working energy of \SI{768.2}{\electronvolt} is indicated. The inset displays XAS on the background where the CoO dots are more isolated (red and thick blue curves), and on the halo close to the nanostructure (green curve).}\label{fig_xas_on_shadow_and_bckgnd}
\end{figure*}

As was shown in the case of three-dimensional samples (with dimensions significantly exceeding the mean free path of photoelectrons, $\sim$\SI{10}{\nano\meter}) lying on a conductive substrate \cite{kimling_peem_coreshell,streubel2012_HemisphericalPyCapsShadowXMCDPEEM,wartelle2015_broadbandSetupForDWmotionInCylNWs,dacol_bp_id}, XMCD-PEEM can be used in transmission or so-called shadow mode. With respect to directly illuminated parts of the substrate, the shadow regions display a reduced intensity according to the magnetization-dependent absorption through the structure. Since dichroism is integrated along the X-ray path through the structure, which may be non-uniform in thickness and/or magnetization configuration, not all the three-dimensional information from the volume is accessible from this projection onto the substrate. Nevertheless, it was shown that the post-processing of simulated micromagnetic configurations allows one to simulate the shadow XMCD-PEEM contrast of a given object with qualitative \cite{dacol_bp_id} and even quantitative agreement \cite{jamet2015_quantitative_xmcd}. This capability is crucial in the case of strongly inhomogeneous textures such as DWs where the shadow contrast pattern may be non-trivial.

In our case, the complexity is reduced as we are interested in assessing the presence of DWs, not their internal structure. Furthermore, the observation of our sample in transmission also allows us to check that in the bulk, the sample is weakly if at all oxidized. Similarly to Electron Energy Loss Spectroscopy (EELS) and with a comparable or somewhat lesser energy resolution \cite{belkhou2015_XRmicroscopyAtHermes,Krivanek2016_10meVresolutionWithSTEM-EELS}, the technique enables to study the characteristic peak of the metal(s) of interest, in our case cobalt, and determine if the bulk material is rather metallic or oxidized based on the absence or presence of a multiplet structure.

\subsection{Imaging and spectroscopic characterization}

Considering the Co$_2$(CO)$_8$ precursor, the bright dots that can be seen around the nanowire consist of cobalt, oxygen and carbon. Their thickness is on the order of \SI{100}{\nano\meter} or less, as observed with SEM. Establishing correspondence between PEEM imaging at the Co L$_3$ edge and SEM imaging is not straightforward because of the interplay between topography and local composition. Cobalt is revealed by the resonant imaging at an absorption edge of this element, while the presence of oxide and/or carbide is indicated by X-ray Absorption Spectroscopy (XAS) on the inset of Fig.\ref{fig_xas_on_shadow_and_bckgnd}: the Co L$_3$ peak displays a multiplet structure incompatible with metallic cobalt \cite{nakajima1999_xmcd_absorptioncoeff}, whereas a similar shape was reported for CoO \cite{deGroot1993_XRabsorptionOfCoOatL3edge} and cobalt carbide \cite{vovan_thesis}. It appears that the dots are probably very similar to one another given the degree of similarity between the spectrum of an ensemble of them (red curve in the inset of Fig.\ref{fig_xas_on_shadow_and_bckgnd}) and that of one given dot (thick blue curve). This peak structure is much more pronounced (as shown by the green curve on the inset of Fig.\ref{fig_xas_on_shadow_and_bckgnd}) closer to the structure, where there seems to be an almost continuous layer. This halo is known \cite{cordoba2010_HighPurityCoNanostructuresFEBID} to originate from secondary electrons created by back-scattered electrons. The spectra on the inset of Fig.\ref{fig_xas_on_shadow_and_bckgnd} were rescaled so that the pre-edge intensities match, in order to allow for comparison.

 On the other hand, the shadow features spectral information about the volume of the nanowire. Fig.\ref{fig_xas_on_shadow_and_bckgnd} illustrates XAS data extracted from the shadow of the nanostructure after subtracting the background averaged over an area of a few square microns close to the shadow's location. The red (blue) curve displays the XAS intensity for the circularly right (left) X-ray polarization. The data is noisy because of the strong background: the signal-to-noise ratio is on the order of a few percents only. Yet, the multiplet structure is much less pronounced if at all present despite the fact that the outer rim of the cobalt rod (where the purity is decreasing) is included in the absorption spectra and thus decreases the averaged Co metallicity. We have therefore a strong indication of a much more metallic material in the volume of the sample, which was protected from oxidation by the Pt coating \cite{pablo-navarro2016_CoreShellFEBIDcoPtNWs}. The arrow indicates the working energy \SI{768.2}{\electronvolt} taken as Co L$_3$. Since the spectra were acquired on a part of the shadow with uniform XMCD contrast [see white segment (4) in Fig.\ref{fig_xmcd_w_profiles}.b)], the difference in intensity should reflect this contrast. This difference is however very small (possibly below noise), nevertheless, this working energy yielded the best quality for XMCD-PEEM images. Imaging was also performed at the top of the peak, at \SI{769.1}{\electronvolt}, as well as \SI{769.85}{\electronvolt} where the difference between the spectra is the largest, but the contrast amplitude at these energies is even lower than at \SI{768.2}{\electronvolt}.

%\begin{figure}[!h]
%\centering
%\includegraphics[width=3.1in]{XAS_CL_CR_minus_bckgnd_v3.png}
%\caption{XAS in the shadow of the nanowire for circular left (blue) and circular right (red) polarization.}
%\label{fig_xas_in_shadow}
%\end{figure}

\subsection{Magnetic state of the wire}

As was indicated in the introduction, before mounting the sample into the microscope, a horizontal field was applied in the direction of the bends with an electromagnet so as to saturate the whole nanowire. The field strength was $\mu_0 H_\mathrm{a}=$~\SI{0.73}{\tesla}. Upon decreasing the strength of the field to zero, three DWs should be nucleated as shown in Fig.\ref{fig_sem_nw_asgrown}.c). Considering the almost cylindrical geometry of the sample as well as its expected soft nature, a field $H_\mathrm{sat}$ equal to half the saturation magnetization should be sufficient to reach saturation. In the case of pure bulk Co, one would have $\mu_0 H_\mathrm{sat}^{(0)}=\mu_0M_\mathrm{s,Co}/2=\ $~\SI{0.89}{\tesla}. However, in our case, we benefit from the angle of about \SI{45}{\degree} of the structure's segments with respect to the field: this reduces the required field by a factor $\sqrt{2}$ and therefore the field needed is: $\mu_0 H_\mathrm{sat}=\ $\SI{0.63}{\tesla}$<\mu_0 H_\mathrm{a}$. Thus, we are sure to have reached saturation without needing to invoke a reduced saturation magnetization with respect to pure bulk cobalt. Moreover, the difference $H_\mathrm{a}-H_\mathrm{sat}$ provides an angular tolerance $\Delta\theta$ on the direction of the applied field of $\Delta\theta=\arccos{(H_\mathrm{sat}/H_\mathrm{a})}=\SI{30}{\degree}$: as long as saturation is reached, tilting magnetization as a whole by less than \SI{45}{\degree} upwards or downwards will not change the end state.

\begin{figure}[!h]
\centering\includegraphics[width=0.5\textwidth]{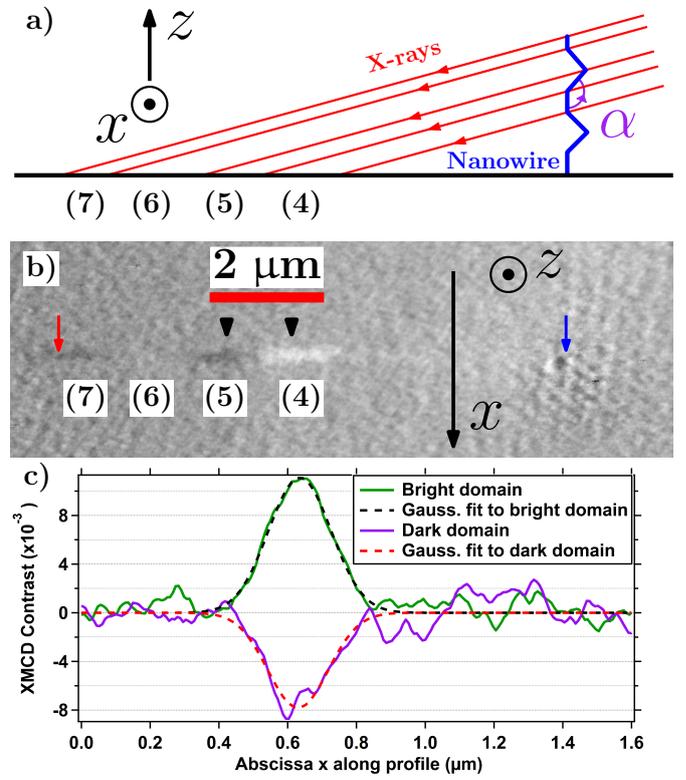}
\caption{a) True-to-scale schematic representation of the X-ray beam falling onto the nanowire; the scale length is as in Fig.\ref{fig_xmcd_w_profiles}.b). The rays passing through junctions between segments delimit the shadow sections corresponding to segments (4) through (7). b) Shadow XMCD-PEEM image of the nanowire with red arrow indicating the tip of its shadow and blue arrow indicating the approximate nanowire position. The domains with uniform contrast correspond to the segments (4), (5) and (7) from Fig.\ref{fig_sem_nw_asgrown}.b). The two black markers above segments (4) and (5) indicate the lateral positions of the contrast profiles in Fig.\ref{fig_xmcd_w_profiles}.c). The transition region between segments (4) and (5) is where a DW was pinned. c) Contrast profiles across the shadow in domains (5) (in purple) and (4) (green), along with their best gaussian fits in dashed lines. The linear gray scale goes from $-2.5$\% to $+2.5$\%. To the right, the noise is slightly stronger where the structure stands because of the field emission.}\label{fig_xmcd_w_profiles}
\end{figure}

A schematic but true-to-scale view of the X-ray beam falling onto the structure is featured in Fig.\ref{fig_xmcd_w_profiles}.a): it illustrates how the subsequent nanowire segments (4) through (7) are projected onto the substrate to form their respective shadow sections. One should note that the nanostructure's possible tilt is not featured. Correspondingly, the XMCD-PEEM image of the structure's shadow shown in Fig.\ref{fig_xmcd_w_profiles}.b), with a red (resp. blue) arrow indicating its tip (resp. the nanowire's position) reproduces these sections, which are vertically aligned with their counterparts in Fig.\ref{fig_xmcd_w_profiles}.a). The structure itself is located on the right on the image, where the noise is stronger due to the field emission causing a strong local reduction of intensity. Away from the shadow, the contrast is zero, indicating that the dots and the halo are not ferromagnetic, as is suggested by the work of Fern\'{a}ndez-Pacheco \emph{et al.} \cite{fernandezpacheco2013_3DnanowiresByMOKE}. The shadow itself features three identifiable domains: a quite faint dark domain at the tip corresponding to segment (7), and further away from the latter another dark domain in segment (5) followed by a bright one in segment (4). First of all, the position of the transition from (4) to (5), about %$\SI{5.6\pm0.3}{\micro\meter}$ 
\SI{5.5\pm0.3}{\micro\meter} away from the structure, is consistent with its SEM-measured height [see Fig.\ref{fig_sem_nw_asgrown}.a)] of \SI{1.5\pm0.1}{\micro\meter}: (\SI{5.5\pm0.3}{\micro\meter})$\times\sin$(\SI{18}{\degree})$=$\SI{1.7\pm0.1}{\micro\meter}. Even though the background emission is strong, the close to 1\% contrast amplitude is clearly above noise in the contrast profiles of Fig.\ref{fig_xmcd_w_profiles}.c). Regarding the magnetic state of the observable parts of the nanowire's shadow, let us first recall that if we name $\varphi$ the angle between the beam's wave vector and magnetization, the XMCD contrast is proportional to $\cos\varphi$. Now, the absence of contrast in segment (6) is due to both the beam's incidence angle (closest to \SI{90}{\degree} among all visible segments) and the presence of a bright dot in this area, as can be seen on the right image in Fig.\ref{fig_cut_shadow}. This dot does not contribute to the dichroic signal because of its non-ferromagnetic nature, but does enhance the background level and therefore reduces the signal-to-noise ratio \cite{jamet2015_quantitative_xmcd}. Similarly, to the right of segment (4), the halo emission is too strong and no magnetic contrast can be retrieved from this part.

The soft nature of the deposited cobalt, along with the constant contrast sign within segments (7), (5) and (4), suggests that magnetization is uniform inside them. As for the change in contrast sign from segment (3) to segment (4), we must take care because its interpretation heavily depends on the angle $\alpha$ (see Fig.\ref{fig_xmcd_w_profiles}) between them. Let us make a though experiment in which one segment is always aligned with the beam, and we change both the magnetization configuration and $\alpha$. Let us consider first that this angle is less than \SI{90}{\degree}, and magnetization follows the wire direction without DW at the bend. In this first case, $\varphi$ changes by \emph{more} than \SI{90}{\degree} from segment (3) to segment (4): as a result, the XMCD contrast being proportional to $\cos\varphi$, it must change sign. On the contrary, if $\alpha$ is still less than \SI{90}{\degree} but a DW sits in the bend, magnetization keeps a projection of the \emph{same} sign onto the beam (the limiting case being $\alpha\rightarrow 0$, where the projections from both segments are equal), and thus the contrast does not change sign. Now, if $\alpha\!>\ $\SI{90}{\degree}, as is the case in our experiment, the configuration where magnetization does not feature a DW between segments (3) and (4) leads to no change in the contrast sign. Indeed, $\varphi$ can change at most by $($\SI{180}{\degree}$-\alpha)\!<\ $\SI{90}{\degree}; the limiting case is a straight wire with uniform magnetization. The last configuration, which we observe, is a change in contrast while $\alpha\!>$ \SI{90}{\degree}, i.e. $\varphi$ changes by more than \SI{90}{\degree}. Following the same reasoning for our experiment, since geometry alone does not allow changes in $\varphi$ of more than \SI{90}{\degree}, we can conclude that magnetization reverses from segment (3) to segment (4): in other words, a DW is pinned in this bend. Thus, we confirm the suitability of our nanowire geometry and material for a controlled creation of DWs.

We tried to compute the expected transversal contrast profile across the structure's shadow and compare its shape with the curves in Fig.\ref{fig_xmcd_w_profiles}.b). The agreement between the data and a gaussian profile $f$ defined by $f(x)=A\exp{[-(x/\Delta)^2]}$ is very good, with fitting results $\Delta^{(\mathrm{exp})}\simeq\ $\SI{120}{\nano\meter} in both cases. It is noteworthy that the actual diameter of the Co rods is about \SI{70}{\nano\meter}}. At first sight, this agreement is surprising because the theoretical profile should be of the form:

\begin{equation}
C(x)=A\tanh\left({\kappa\sqrt{1-\frac{x^2}{d^2}}}\right)
\end{equation} \\ where $\kappa$ is proportional to the difference $\delta\mu$ in linear absorption coefficients of X-rays in Co (for magnetization parallel and anti-parallel to the beam at the Co L$_3$ edge) and the travelled distance through the structure. Here, $d$ is the diameter of the nanowire. Considering the length scales as well as the order of magnitude of $\delta\mu\simeq\ $\SI{0.034}{\nano\meter\tothe{-1}} \cite{nakajima1999_xmcd_absorptioncoeff}, one would expect a profile much closer to a square function than to a gaussian since we are in the strong absorption regime (except close to the edges), where dichroism should saturate. The origin of this shape is partly wave optics: the nanowire diffracts the incoming beam, which leads to an image onto the substrate with different lateral dimensions. Fringe patterns due to Fresnel diffraction on 3D objects have already been reported \cite{jamet2015_quantitative_xmcd} in XMCD-PEEM experiments. In our case, we are dealing with Fresnel diffraction as well, and a rough numerical computation of the diffraction pattern (not shown here) leads to a satisfactory agreement in terms of profile shape. However, taking into account the instrumental resolution of ca. \SI{30}{\nano\meter} as well as the beam's divergence of ca. \SI{0.008}{\milli\radian} leads to an estimate of the width $\Delta^{(\mathrm{theo})}\simeq\ $\SI{80}{\nano\meter}, in rather poor agreement with $\Delta^{(\mathrm{exp})}$. 

A plausible source of broadening for the profile is vibrations of the vertical nanowire caused by fluctuations in the microscope's high voltage and the field emission process. Since the structure is not entirely metallic, variations in the electric field to which it is subjected would lead to random forces exerted upon its tip. Considering how our integration time for image acquisition is on the order of a second, voltage noise of frequencies above a few hertz would lead to a blurred and wider shadow. This is consistent with what can be observed in Fig.\ref{xmcd-peem_imaging}: the shadow appears wider close to its tip with respect to sections closer to the nanostructure's foot. Furthermore, such vibrations account for the profile shape, the absence of diffraction fringes, and contribute to decreasing the shadow contrast as it is spread over a wider area. Regarding the DW, the absence of clear features in the corresponding shadow region is also consistent with them being averaged out because of vibrations.

%\subsection{?}

\section{Conclusion}
In summary, we have fabricated vertical, Pt-coated Co nanowires featuring \SI{90}{\degree} bends with FEBID and successfully investigated one such structure using shadow XMCD-PEEM. The bulk material quality was checked by XAS in spite of strong background emission from Co-based dots and a continuous halo surrounding the structure. Such an impediment to observations could be suppressed by deposition of a thick enough metallic layer to prevent the extraction of photoelectrons from Co in the substrate, say \SI{10}{\nano\meter} of gold or platinum. In this frame, a much larger extent of the structures' shadow should be revealed up to the close vicinity of the structure, and a significantly larger signal-to-noise ratio achieved: the $\sim1\%$ contrast amplitude is by far not an upper limit. Moreover, the Co content and associated magnetization could be increased by post-growth annealing strategies, thus enhancing the magnetic contrast \cite{begun2015_PurificationOfFEBIDcobaltForNanostructures,dosSantos2016_AnnealingOfFEBIDcobaltForElectricalTuning}. A more severe limit is the difficult (if at all possible) direct imaging caused by the field emission. The latter has supposedly contributed to distorting the nanostructure, but the majority of the structures that were glimpsed in the experiments were still standing, which testifies of their robustness despite their non-conventional geometry. Thanks to the latter and the application of an in-plane field, we were able to conclude from the shadow XMCD-PEEM contrast that at least one DW was present in the structure. Our interpretation of the contrast is supported by our theoretical simulations of the XMCD pattern, which highlights the role of Fresnel diffraction in such investigations \cite{jamet2015_quantitative_xmcd}. It appears that the biggest issue for the imaging far from the structure lies in the vibrations undergone by the latter due to voltage noise and the field emission in the microscope. This would first call for confirmation through imaging with lower integration times. Then, using either magnetic field gradients or piezo actuators, one could set the structure tip into periodic motion; in this frame, proper synchronization and low integration times would allow to acquire images with reduced influence of the vibration while keeping the signal to noise ratio sufficiently high. Care would be needed in the case of a magnetic field so that the latter does not disturb the micromagnetic configuration of the nanowire, however, there also is a challenge in producing sufficiently strong field gradients in the microscope environment. On the other hand, the piezo actuators would provide a more reliable and purely mechanical source of oscillations, with the added value of higher bandwidth. This may seem like an impediment to recovering the full information from the material, but in this frame, the use of tomographic methods and the inclusion of diffraction in the simulations of shadow XMCD-PEEM contrast patterns should allow the resolution of non-uniform magnetization configurations such as DWs. If however the nanowire is in a self-oscillating regime due to an electric-field-gradient-induced mechanical instability \cite{Lazarus2010_SelfOscillationInNEMS}, the difficulty is increased as the tip motion occurs at the fundamental vibration mode of the nanowire, which a rough estimate places at a frequency $\nu_1$ on the order of tens of megahertz.

Investigating such nanostructures in a horizontal configuration would be easier. One could think of micromanipulating these onto two laterally separated FEBID-fabricated supports, as a manner of bridge between elevated posts. Not only would this avoid the strong field emission effects that we encountered, it would also allow reduce the structure's vibrations as it could be clamped in place at both its extremities. In this perspective, all the structural engineering described in this report could still be put to use as the structure would be plucked off after its fabrication. 

\section*{Acknowledgements}

The authors would like to thank Nicolas Rougemaille, Johann Coraux, Olivier Arcizet, Benjamin Pigeau and Jean-Christophe Toussaint for fruitful discussions. This work was supported by Spanish Ministry of Economy and Competitivity through projects No. MAT2014-51982C2-1-R, MAT2014-51982C2-2-R and MAT2015-69725-REDT, including FEDER funds, and by the Aragon Regional Government (Construyendo Europa desde Arag\'{o}n) through project E26, with FEDER funding. Javier Pablo-Navarro grant is funded by the Ayuda para Contratos Predoctorales para la Formaci\'{o}n de Doctores, Convocatoria Res. 05/06/15 (BOE 12/06/15) of the Secretar\'{i}a de Estado de Investigaci\'{o}n, Desarrollo e Innovaci\'{o}n in the Subprograma Estatal de Formaci\'{o}n of the Spanish Ministry of Economy and Competitiveness (MINECO) with the participation of the European Social Fund. Discussions with Dr. Luis Serrano-Ramón and Dr. Amalio Fern\'{a}ndez-Pacheco about the growth of the Co FEBID nanowires are warmly acknowledged. Michal Sta\v{n}o acknowledges grant from the Laboratoire d'excellence LANEF in Grenoble (ANR-10-LABX-51-01).

\bibliographystyle{unsrtnat}
\bibliography{bibliography}

\end{document}